\documentclass[draft]{agujournal2019}
\usepackage{url} 
\usepackage[inline]{trackchanges} 
\usepackage{soul}
\usepackage{mathtools}
\draftfalse

\journalname{Geophysical Research Letters}

\begin{document}

\title{Earth's Alfv\'{e}n Wings: Unveiling Dynamic Variations of Field-line Topologies with Electron Distributions}

\authors{Harsha Gurram\affil{1,2}, Jason R. Shuster\affil{3}, Li-Jen Chen\affil{2}, Hiroshi Hasegawa\affil{4}, Richard E. Denton\affil{4}, 
 Brandon L. Burkholder\affil{2,5}, Jason Beedle\affil{3}, Daniel J. Gershman\affil{2}, James Burch\affil{7}}
\affiliation{1}{Department of Astronomy, University of Maryland College Park, College Park, Maryland, USA. }
\affiliation{2}{NASA Goddard Space Flight Center, Greenbelt, Maryland, USA.}
\affiliation{3}{Space Science Center, University of New Hampshire, Durham, New Hampshire, USA.}
\affiliation{4}{Institute of Space and Astronautical Science, JAXA, Sagamihara, Japan}
\affiliation{5}{Department of Physics and Astronomy, Dartmouth College, Hanover, New Hampshire, USA.}
\affiliation{6}{University of Maryland Baltimore, Baltimore, Maryland, USA.}
\affiliation{7}{Southwest Research Institute, San Antonio, TX, USA.}

\correspondingauthor{Harsha Gurram}{hgurram@umd.edu}

\begin{keypoints}
\item First characterization of electron distributions within sub-Alfv\'{e}nic solar wind and Alfv\'{e}n wing flux tubes during a CME.  
\item Electron distribution functions within newly closed field regions formed by dual wing reconnection exhibit four distinct populations.
\item Electron distribution functions suggest dynamic reconnection turning on and off.
\end{keypoints}
\newcommand{\scr}[1]{_{\mbox{\protect\scriptsize #1}} }

\begin{abstract}
The magnetic cloud (MC) of the Coronal Mass Ejection on April 24, 2023, contains sub-Alfv\'{e}nic solar wind, transforming Earth's magnetosphere from conventional bow-shock magnetotail configuration to Alfv\'{e}n wings. Utilizing measurements from the Magnetosphere Multiscale (MMS) mission, we present for the first time electron distribution signatures as the spacecraft traverses through various magnetic topologies during this transformation. Specifically, we characterize electrons inside the sub-Alfv\'{e}nic MC, on the dawn-dusk wing field lines and on the closed field lines. The signatures include strahl electrons in MC regions and energetic keV electrons streaming along the dawn and dusk wing field lines. We demonstrate the distribution signatures of dual wing reconnection, defined as reconnection between dawn-dusk Alfv\'{e}n wing field lines and the IMF. These signatures include four electron populations comprised of partially-depleted MC electrons and bi-directional energetic electrons with variations in energy and pitch-angle. The distributions reveal evidence of bursty magnetic reconnection under northward IMF.

\end{abstract}

\section*{Plain Language Summary}
Similar to how a bow shock forms when a supersonic flow hits an object to slow it down, Earth's bow shock is formed to slow down the incoming solar wind. However, during events like the coronal mass ejection on April 24, 2023, the solar wind density drops so low that the bow shock doesn't form, causing Earth's magnetosphere to change into a new shape that includes "Alfv\'{e}n wings". In this new configuration, Earth's magnetotail splits into two wings: the dawn wing and the dusk wing. Such occurrences are rare at Earth but are commonly observed in the magnetosphere of Ganymede, Jupiter's moon, and other inner planets. Due to limited instrumentation, no studies have explored the transformation into wings from the perspective of electrons. We present the first study of this kind, utilizing the high-precision suite on the Magnetosphere Multiscale (MMS) mission to unveil the dynamic variations in magnetic topology by characterizing electrons in the various topologies. We present the electron distribution signatures of IMF reconnection with both the dawn-wing and dusk-wing field lines (termed dual-wing reconnection), and estimate the age and intensity of the reconnection sites.

\section{Introduction}

The solar wind expands continuously through space, reaching super-Alfv\'{e}nic speeds as it approaches Earth, typically with a Mach number($M_A$) around 8. Occasionally during Coronal Mass Ejection (CME) events, however, the magnetic field within the CME intensifies, and the plasma density decreases significantly, resulting in sub-Alfv\'{e}nic conditions (\textit{i.e.} $M_A$ less than one) \cite{Schunk_Nagy_2000}. Under normal circumstances, a bow shock forms in front of Earth to decelerate the super-Alfv\'{e}nic solar wind. However, when the sub-Alfv\'{e}nic solar wind from CMEs reaches Earth, a bow shock does not develop. Instead, the magnetosphere forms a pair of tube-like structures called Alfv\'{e}n wings, which serve to slow down the plasma \cite{chane_2012,ridley_alfven_2007}. These wings extend at an angle to the prevailing magnetic field and reach far into space, illustrating a critical aspect of how Earth's magnetic environment interacts with variations in solar wind conditions \cite{ridley_alfven_2007}.

On 23th April 2023, a CME flux rope erupted on the surface of the sun, driving a magnetic cloud (MC). On April 24th 2023, the CME MC with sub-Alfv\'{e}nic solar wind passed over the Magnetospheric Multiscale (MMS) \cite{burch2016} spacecraft which were located at the dawn flank of the magnetopause during its inbound trajectory. When the sub-Alfv\'{e}nic MC encounters Earth it transforms the typical unified magnetosphere i.e bow-shock and magnetotail, into an Alfv\'{e}n wing formation. The formation of Alfv\'{e}n wings during this CME has been confirmed with global magnetosphere simulations \cite{burkholder_global_2024,chen_interplanetary_2024}. 
\begin{figure}
\includegraphics[width=\textwidth]{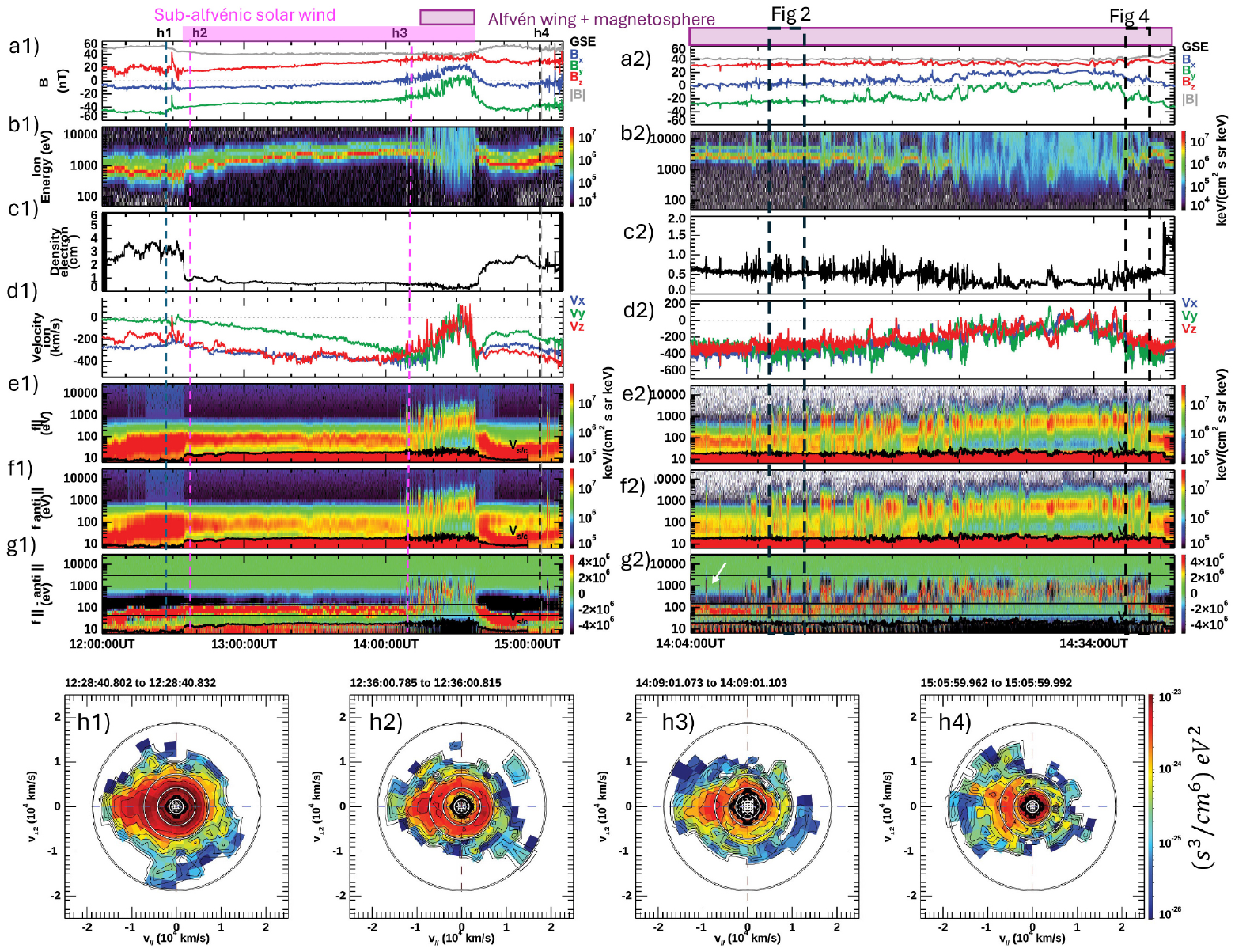}
    \caption{MMS1 observations of the the CME magnetic cloud and sub-Alfv\'{e}nic solar wind (highlighted with purple bar) where the magnetosphere transforms into Alfv\'{e}n wings. The data products used in a1-f1) are in the fast survey mode and those in a2-g2) are the in burst mode data. a1-a2) Magnetic field components and $|B|$ , b1-b2) ion energy spectrogram, c1-c2) electron density, $n_e$, d1-d2) ion velocities, $v_i$, e1-e2) parallel electron energy flux, $f_{||}$, f1-f2) anti-parallel electron energy flux, $f_{\scr{anti}-||}$ and g1-g2) difference between parallel and anti-parallel energy flux, $f_{||}-f_{\scr{anti}-||}$. $B$ and $v_i$ are in GSE coordinates. Solid  lines in g1-2 and circles in the 2D eVDF mark the energy levels at spacecraft potential ($V_{s/c}$), 45eV, 150eV, and 3keV. Electron velocity distribution functions for h1) shocked solar wind plasma, h2) unshocked shocked solar wind plasma h3) CME MC plasma and h4) solar wind plasma during the magnetosphere recovery phase from Alfv\'{e}n wing configuration.   }
\label{fig:eventoverview}
\end{figure}

MMS encounters this sub-Alfv\'{e}nic CME MC plasma at around 12:30UT \cite{chen_earths_2024,beedle_field-aligned_2024}.
Fig.~\ref{fig:eventoverview} provides an overview of the interval where the Earth's magnetosphere transforms into the Alfv\'{e}n's wings. Fig.~\ref{fig:eventoverview} a1-g1 shows the fast survey data and the data products shown in a2-g2 are in the burst mode. This study analyzes data from Fast Plasma Investigation (FPI)\cite{FPI} and Fluxgate Magnetometer (FGM) \cite{FGM} instruments, which provided observations of plasma properties and magnetic field conditions from the MMS1 spacecraft. MMS1 measures the $\sim 2$ hours of sub-Alfv\'{e}nic interval $\sim$ 13:00 - 15:00 UT highlighted with a magenta bar. During this interval the density drops from super-Alfv\'{e}nic to sub-Alfv\'{e}nic solar wind $\sim 5$ cm$^{-3}$ to $\sim 0.5$ cm$^{-3}$ and the interplanetary magnetic field (IMF) has a strong $B_y$ component. The ion omni-directional energy flux shows a narrow beam of $1$keV proton and alpha particles. From 14:00 to 14:40UT MMS1 encounters dawn side Alfv\'{e}n wing and closed fieldline region- magnetosphere (highlighted by purple bar and shown in Fig.~\ref{fig:eventoverview} a2-g2), which is confirmed by magnetic field components and the density is roughly $~0.25$cm$^{-3}$ as shown in Fig.~\ref{fig:eventoverview}b2. Throughout this interval, during encounters with Alfv\'{e}n wing filaments we see density intrusions which are accompanied with fluctuations in magnetic field and increase in ion energy \cite{chen_earths_2024, beedle_field-aligned_2024}.

To visualize the topology of wing and newly formed magnetosphere from the perspective of electrons we analyze the parallel and anti-parallel electron energy fluxes. Fig.~\ref{fig:eventoverview} panels d1-d2 and e1-e2 illustrate parallel and anti-parallel electron fluxes for the three hour interval and the Alfv\'{e}n wing interval respectively. The energy fluxes in parallel and anti-parallel directions are not identical; certain areas exhibit elevated parallel electron fluxes, while others show increased anti-parallel electron fluxes. To visualize this variation, we calculate 
\begin{equation}
    f_{||} - f_{anti-||},
\end{equation}

the difference between the parallel and anti-parallel electron fluxes shown in Fig.~\ref{fig:eventoverview}g1. It demonstrates this difference for shocked, unshocked and Alfv\'{e}n wing intervals. We identify the solar wind plasma, exhibiting a notable bipolar characteristic, delineated by energy bands colored blue (around 200 eV) and red (around 50 eV). The blue band is a result of the strong flux of the strahl electrons emanating from the solar corona with a typical energy of 272 eV at 1 AU \cite{strahl}. The distinction between the unshocked and shocked solar wind is the increased thickness of these bipolar bands in the shocked state, corresponding to change in density. Fig.~\ref{fig:eventoverview}h1-~\ref{fig:eventoverview}h4 shows 2D electron velocity distribution functions (eVDFs) in the $v_{||}-v_{\bot1}$ plane where $v_{\bot1}$ is in the direction of $\mathbf{v_e} \times \mathbf{B}$, where $\mathbf{v_e}$ is the electron velocity vector and $\mathbf{B}$ is magnetic field vector. These distributions represent energy weighted VDFs. The plotted quantity is proportional to the differential energy flux \textit{i.e.} $f(v_{||},v_{\bot 1},v_{\bot 2})*v^4$ to emphasize the features at high velocities.

Fig.~\ref{fig:eventoverview}h1 and h2 shows the 2D eVDF for the shocked and unshocked solar wind near the beginning of sub-Alfv\'{e}nic solar wind interval. Both eVDFs exhibit a narrow beam in $v_{||}<0$ until few hundred eVs ($\sim 200$eV) but the eVDF corresponding to unshocked solar wind has a lower phase space density. Fig.~\ref{fig:eventoverview}h3 illustrates the eVDF inside the sub-Alfv\'{enic} CME MC plasma. The eVDF exhibits a narrow strahl compared to the unshocked solar wind (Fig.~\ref{fig:eventoverview}h2) with further drop in the phase space density. With the help of this eVDF and the corresponding bipolar signature in $f_{||}-f_{anti-||}$ we identify and characterize electrons at various other topologies. Fig.~\ref{fig:eventoverview}h4 shows the eVDF of the solar wind during the magnetosphere recovery phase from Alfv\'{e}n wing configuration that occurs from 15:05-15:15UT \cite{beedle_field-aligned_2024}. The solar wind electrons exhibit: 1) a broader strahl that extends $<200$eV  and 2) a drop (to $<50$eV) in the $||$ energy flux when compared to CME MC plasma as well as the unshocked and shocked solar wind from the earlier times.

At 14:04UT, as MMS encounters the first Alfv\'{e}n wing filament the characteristic bipolar signature of the solar wind extending upto 200eV  disappears, revealing new features that extend to approximately 3keV (indicated by arrow in Fig.~\ref{fig:eventoverview}g2). These are attributed to Alfv\'{e}n wing filaments (highlighted in Fig.\ref{fig:alfvenfilament}) and a closed-field line region (highlighted in Fig.\ref{fig:DualReconnection}). Subsequent sections will delve deeper into the Alfv\'{e}n wing filaments and the closed field line regions through analysis of 2D distribution functions to elucidate the magnetic topology.

\section{Characterizing electrons distributions at different topologies}
\begin{figure}
\centering
\includegraphics[width=1\textwidth]{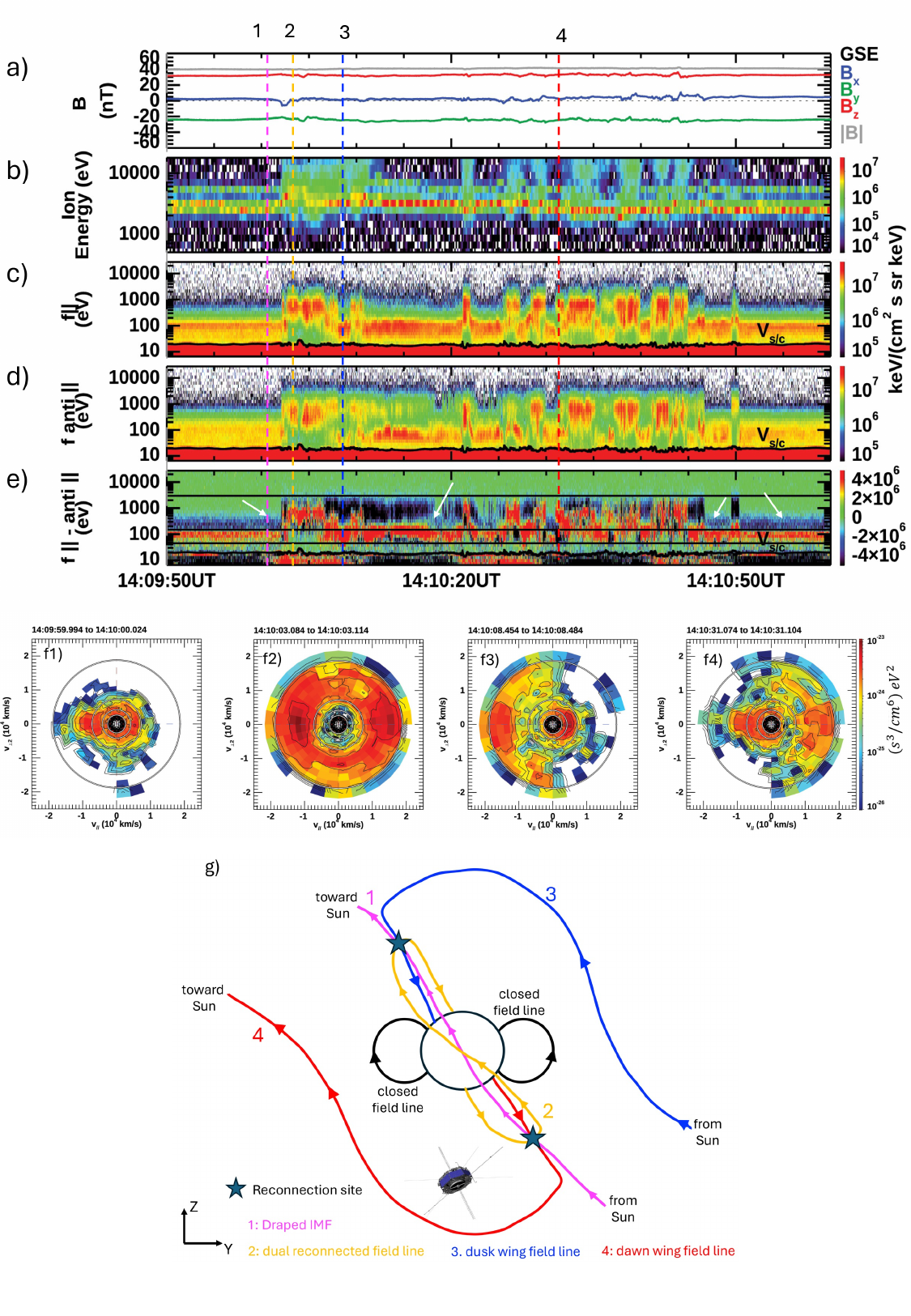}
\caption{Signatures in electron DFs as MMS encounters dawn, dusk and closed field lines. (a) Magnetic field, (b) omnidirectional ion energy flux, (c) parallel energy flux, $f_{||}$, (d) anti-parallel energy flux, $f_{\scr{anti}-||}$ and (e) $f_{||}-f_{\scr{anti}-||}$. The color of the each marked line represents the distinct topology: magenta-MC plasma, yellow-closed field line region, blue-dusk wing field line and red -dawn wing field line. (f1-f4) Energy flux distribution function slices representing different topologies. g) Cartoon showing all marked topologies wherein dual reconnected fieldline forms due to the reconnection between draped IMF and dawn-dusk fieldline. }
\label{fig:alfvenfilament}
\end{figure}

Fig.~\ref{fig:alfvenfilament} provides a zoom-in view of the interval 14:09:50-14:11:00 which features encounter with Alfv\'{e}n wing filaments. Fig.~\ref{fig:alfvenfilament}a shows the magnetic field components with a strong $B_y$ corresponding to the CME MC field. The magnetic field components $B_x(\sim 0)$, $B_y(\sim -20nT)$ and $B_z (\sim 30nT)$ stay constant for most of the interval. During this interval MMS encounters all possible magnetic topologies.
In this section, we characterize the electron distributions corresponding to these different topologies. 

From 14:09:50UT to $\sim$14:10UT MMS sees MC/unshocked solar wind plasma. The corresponding distribution is illustrated in Fig.~\ref{fig:alfvenfilament}f1 which reveals a pronounced, narrow beam of strahl electrons in $v_{||}<0$ (marked by magenta dashed line in Fig.~\ref{fig:alfvenfilament}a-~\ref{fig:alfvenfilament}e). Around $\sim$14:10:03UT, MMS records roughly balanced energetic electron fluxes (marked as a yellow dashed line in Fig.~\ref{fig:alfvenfilament}a-e) and yellow field line in Fig.~\ref{fig:alfvenfilament}g.  The corresponding eVDF is illustrated in Fig.~\ref{fig:alfvenfilament} f2 which illustrates bi-directional keV electrons. This corresponds to closed-field lines due to the reconnection between draped IMF field line with dawn-dusk fieldlines, a process we term dual-wing reconnection. 

We dedicate the next section to discussing this in further depth. The variations in the flux intensities could indicate the strength of reconnection that produces dusk and dawn wing fieldline or amount of time since reconnection ceased. eVDF in Fig.~\ref{fig:alfvenfilament}f3 suggests 
a slightly stronger reconnection corresponding to the dusk wing field line. From $\sim$14:10:03UT to $\sim$ 14:10:07UT, MMS observes these closed field lines. The corresponding signature in $f_{||}-f_{anti-||}$ has no discernible electron anisotropy i.e. no bipolar signature. 


At 14:10:08.454UT, MMS sees dusk-wing field line (marked by a blue dashed line in Fig.~\ref{fig:alfvenfilament}a-e and blue field line in Fig.~\ref{fig:alfvenfilament}g) and the corresponding eVDF is shown in Fig.~\ref{fig:alfvenfilament}f3. The anti-parallel energy flux indicates an energetic electron population at $\sim1$keV, whereas the parallel energy flux exhibits features inside the MC (as in Fig.~\ref{fig:alfvenfilament}f1). The source of these anti-parallel energetic electrons is either Earth or the interactions between the MC and the magnetosphere, facilitated by magnetic reconnection. However, the characteristics of the magnetic field components, illustrated in panel Fig.~\ref{fig:alfvenfilament}a, resemble those of the MC or solar wind, thereby excluding the possibility of a magnetospheric origin. This observation substantiates that the eVDF like Fig.~\ref{fig:alfvenfilament}f3 are consistent with dusk wing flux tubes, generated by the IMF reconnecting with closed field lines northward of the MMS. The anti-parallel energetic electrons have stronger intensity and are more spread-out in pitch angles. This corresponds to an "aged" dusk-wing field line as greater spread in pitch angles implies the distribution has been on the field line for a long time. Consequently, the dusk wing field lines are characterized by energetic anti$-||$ electrons +  $||$ MC electron population. 

MMS sees often these "aged" dusk-wing field lines during the next $\sim$ 12 seconds, indicated by anti-$||$ electrons extending to 3 keV with concurrent parallel electrons at approximately the MC electron energies, which is suggested by the bipolar feature in $f_{||}-f_{\scr{anti}-||}$. The bipolar feature comprises of blue energy band extends from a few hundred eV to 3keV and the red energy band from few tens of eV to around few hundred eV. At around 14:10:20UT MMS again observes MC plasma (indicated by white arrow) and closed field lines topologies. 

The eVDF in Fig.~\ref{fig:alfvenfilament}f4 exhibits field-aligned energetic electrons and a MC-like population in the anti-parallel direction. The energy fluxes are again imbalanced, but this time the higher energy flux occurs in the parallel direction. This is consistent with dawn wing field lines (red field line in Fig.~\ref{fig:alfvenfilament}g) generated by IMF reconnecting with closed field lines southward/duskward of MMS (energetic $||$ electrons +  anti-$||$ MC electron population).   For the remainder of the interval, MMS predominantly detects closed field lines, dusk-wing flux tubes, and MC plasma. These observations can be deduced from the corresponding signatures in Fig.~\ref{fig:alfvenfilament}e as described in the above for Figs.~\ref{fig:alfvenfilament} f2, f3 and f1 respectively. Concurrently, despite being slowed down and variably heated, ions exhibit characteristics of the MC ion population \cite{chen_earths_2024} throughout this interval. 
\begin{figure}
\includegraphics[width=1\textwidth]{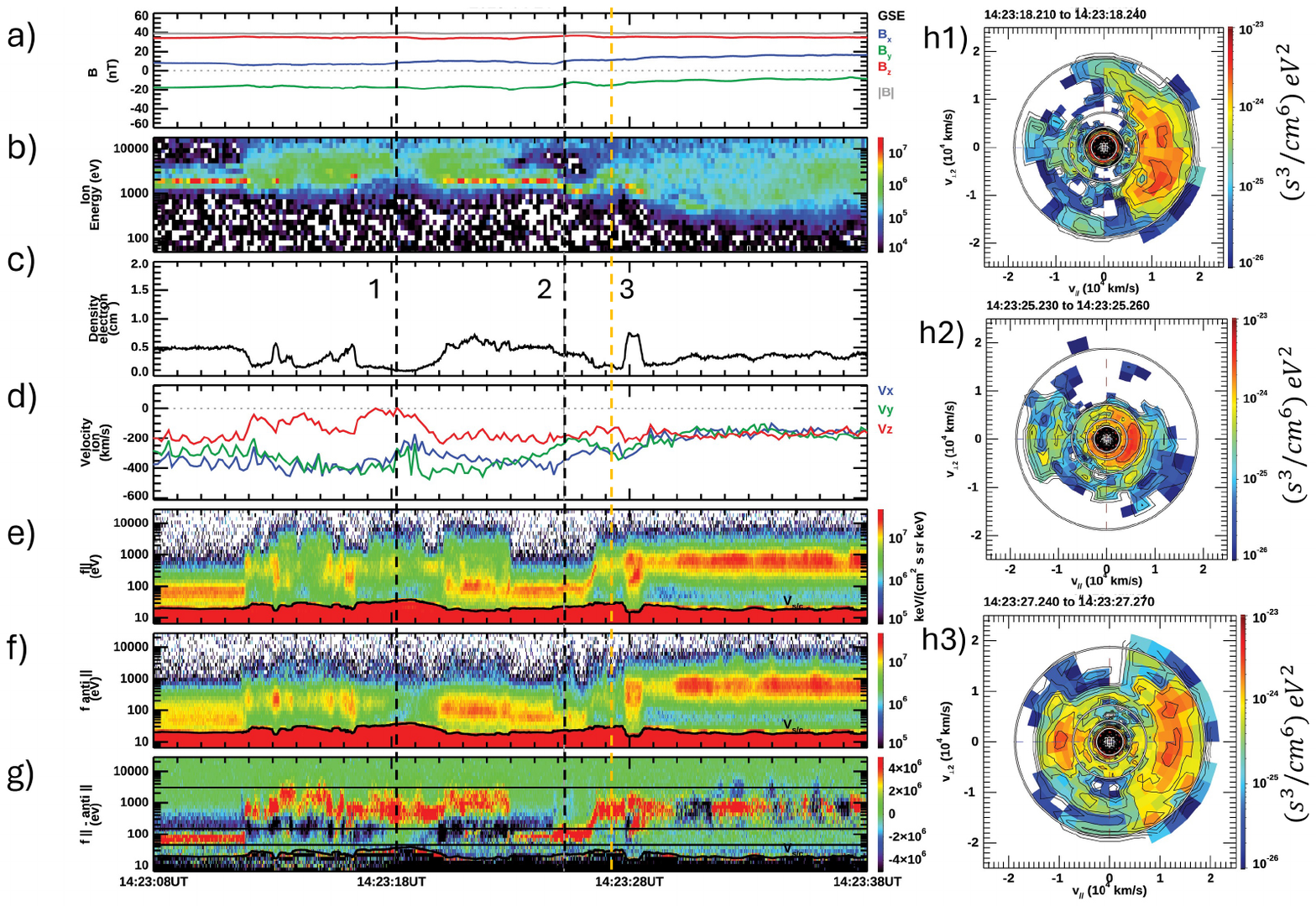}
\caption{Signatures in electron VDFs when MMS is at the entry edge into the closed field line region. (a) Magnetic field, (b) omnidirectional ion energy flux, (c) electron density, (d) parallel energy flux, $f_{||}$, (e) anti-parallel energy flux, $f_{\scr{anti}-||}$ and (f) $f_{||}-f_{\scr{anti}-||}$. (h1-h3) 2D eVDF at  various topologies.}
\label{fig:strahlmiss}
\end{figure}

At the entry into the closed field line region \textit{i.e} 14:23-14:25UT MMS observes eVDFs consistent with Alfv\'{e}n wing filaments and closed field line regions. Fig.~\ref{fig:strahlmiss} shows a 20s interval at the entry edge into freshly-closed regions. MMS observed MC plasma $\sim$ 14:23:12UT and closed field regions until 14:23:18UT. From 14:23:18.110UT MMS encounter new fieldlines lasting for about $\sim2$ seconds. In this interval the magnetic field components are MC-like however the omni ion energy flux shows no MC-like ion population and the electron density drops to $0.25$cm$^{-3}$ - which is more magnetosphere like (marked by first blacked dashed line in Fig.~\ref{fig:strahlmiss}a-~\ref{fig:strahlmiss}g). The parallel and anti-parallel energy fluxes exhibit significant depletion at low energies ($<200$eV), and their difference shows a unipolar red band around 1keV.
The corresponding eVDF is shown in Fig.~\ref{fig:strahlmiss}h1, which confirms the absence of MC electrons- \textit{i.e.} no strahl electrons in ($v_{||}<0$) as well as absence of $||$ electrons around $100$eV. This eVDF likely represents a closed-field topology with the anti-parallel MC electrons entirely depleted and no source of new electrons from the Earth. One possible scenario to account for the lack of anti-parallel electrons is to have dual-wing reconnection where the reconnection site northward of MMS has become dormant. The observation alone cannot exclude the possibility that the anti-parallel part of the field lines are connected to a special solar corona region with little plasma source.

Fig.~\ref{fig:strahlmiss}h2 illustrates the eVDF at 14:23:25.230UT, where MMS observes MC-like magnetic field, MC-like ions and MC-like parallel energy flux. However, the strahl component of the MC plasma is absent in $f_{\scr{anti}-||}$ and in the eVDF. This topology could potentially be interpreted as either: (1) MC plasma with the field line no longer connected to the solar corona, or (2) a dusk wing field line, given that the parallel features resemble those in Fig. \ref{fig:alfvenfilament}f3, with reconnection north of the MMS having ceased, thereby lacking signatures of energetic electrons. Based on the established characterizations, this topology cannot be determined conclusively and remains an open question. At 14:24:27.240 UT, MMS observes features similar to those at earlier time marked as 1 in Fig.\ref{fig:strahlmiss}: 1) no MC-like ion beam, 2) no MC electrons, 3) higher parallel electron flux than anti-parallel flux, manifesting as a red band in $f_{||}-f_{\scr{anti}-||}$. The difference lies in the anti-parallel flux; at Fig.~\ref{fig:strahlmiss}g3) there is a slightly higher $f_{\scr{anti}-||}$ in the energy range $50-100$eV (faint blue band in this energy range in g). This slight difference significantly impacts the signatures of $v_{||}<0$ in the VDF, accounting  for different topologies whether the reconnection northward of MMS is active or dormant or if the field line is connected to the solar corona with little source. Thus, while various topologies exhibit distinct signatures in $f_{||}$ , $f_{\scr{anti}-||}$ and $f_{||}-f_{\scr{anti}-||}$, a detailed investigation of electron distributions is essential for conclusive determination.

One interesting aspect to consider during the encounter with the Alfv\'{e}n wings is the variation in dispersion of  parallel and anti-parallel electron energy fluxes. For example, close to the blue dashed line in Fig.\ref{fig:alfvenfilament},  the electron energy fluxes exhibit distinct characteristics. The parallel electron flux displays dispersion, in contrast to the anti-parallel electrons which exhibit multiple electron populations and no dispersion. Between 14:10:25-14:10:30UT there are further variations, with the parallel flux dispersing to higher energies and the anti-parallel flux dispersing towards lower energies. These variations persist throughout the sub-Alfv\'{e}nic interval, as illustrated in Fig.\ref{fig:strahlmiss} (between 2 and 3 dashed lines). The discrepancies in flux dispersion warrant further investigation and may offer significant scientific insights into the mechanisms of energy transport across different topologies.

\section{Electron distributions in freshly-closed field line regions resulting from dual wing reconnection}

The magnetic reconnection between the IMF and the Earth's magnetospheric field line results in an open field line topology where the magnetic field line has one end at the Earth and the other at the Sun. This reconnection results in heated electrons streaming on open-field lines. Analogous to dual-lobe reconnection where magnetic reconnection on lobe field lines poleward of both cusps can produce a closed field topology \cite{Onsager_2001,Lavraud_2005a,Lavraud_2005b} and multiple ion populations in the low-latitude boundary layer \cite{Bavassano_2006}, here the closed field-line regions are formed due to reconnection between the IMF and dusk- as well as dawn-wing field lines, referred as dual-wing reconnection. The closed field line regions correspond to intervals with roughly balanced bi-directional keV electrons. Parallel and anti-parallel energetic electrons stream along these field lines from reconnection sites located southward and northward of MMS, respectively. We can infer whether the dual-reconnected field lines are younger or older by examining the pitch angle distribution of the keV electrons and the intensity of the remaining MC electrons in the eVDFs. Greater depletion of MC electrons suggests that the field line is older. Conversely, if the lower energy population is similar to the MC electron population, the field line is likely so young that the MC electrons have not had time to be depleted.

\begin{figure}
\centering
\includegraphics[width=1.1\textwidth]{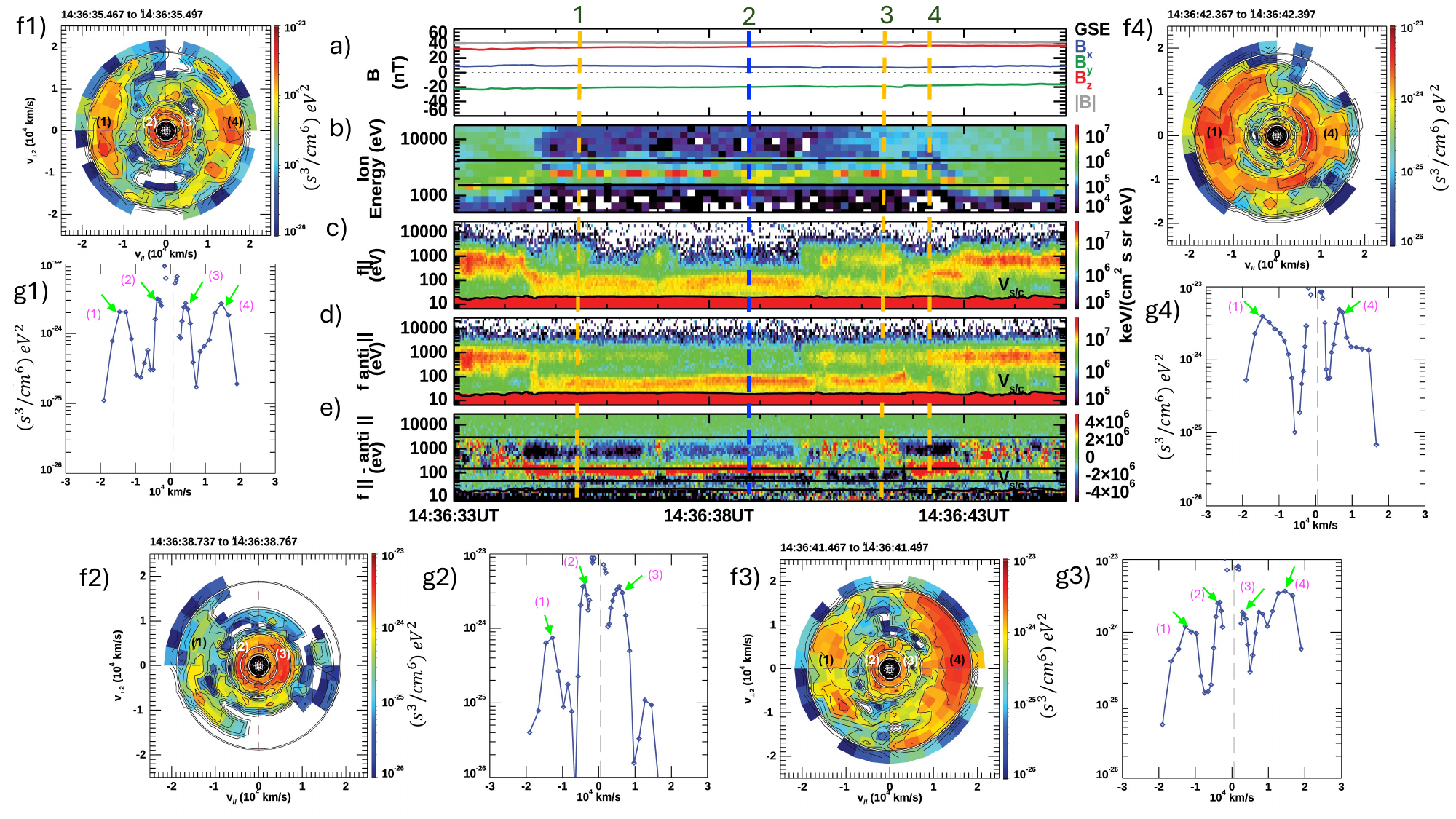}
\caption{Signatures in electron VDFs as MMS encounters various magentic  topologies between 14:36:33 -14:36:45 UT. (a) Magnetic field, (b) omnidirectional ion energy flux, (c) $f_{||}$, (d) $f_{\scr{anti}-||}$ and (e)$f_{||}-f_{\scr{anti}-||}$. f1-f4) 2D electron VDFs.  g1-g4) 1D cuts along $v_{||}$ on these field lines.}
\label{fig:DualReconnection}
\end{figure}


From 14:25-14:35UT MMS primarily encounters closed field line topology, as the electrons are bi-directional and more energetic than MC electrons, and additionally, the magnetic field components are similar to those in the magnetosphere at 17:00UT. Furthermore, electrons are hot and the density ($0.25$cm$^{-3}$) is lower than the MC density ($0.5$cm$^{-3}$) which is accompanied by enhanced fluxes of 60-150 keV electrons with pitch angles that peak at approximately 90 degrees \cite{chen_earths_2024}. MMS exits this magnetosphere-like region around $\sim$14:35UT and encounters mixed field line topologies. Fig.~\ref{fig:DualReconnection} highlights a 12s interval to show on regions with both balanced electron fluxes and regions with slightly larger anti-parallel fluxes.

In the interval shown in Fig.\ref{fig:DualReconnection}, we note the co-existence of keV electron populations with lower energy populations both parallel and anti-parallel.  Fig.~\ref{fig:DualReconnection}g1-f1 illustrates the energy distribution function along with a one-dimensional cut in the parallel direction. This cut distinctly identifies four electron populations two each aligned parallel and anti-parallel. Populations labeled (2) and (3) are characterized by energies in the range of a few hundred electron volts, whereas those labeled (1) and (4) align with approximately 1 keV electrons. Populations (2) and (3) represent the partially depleted MC populations. The presence of bi-directional keV electrons indicates reconnection between dusk and dawn-wing field lines. The presence of MC-like electrons along with low pitch-angle spread in energetic bi-directional electrons are consistent with a configuration associated with newly/freshly reconnected closed-field line topologies.

The distribution at 14:36:38.737UT, as depicted in Fig.\ref{fig:DualReconnection}g2-f2, reveals three distinct populations, corresponding to a tripolar signature in $f_{||}-f_{\scr{anti}-||}$. In comparison to the flux observed at 14:36:34.100UT, the electron energy flux exhibits characteristics more akin to those seen earlier, specifically the MC-like pattern evident in the $||$ flux shown in Fig.\ref{fig:alfvenfilament}f3.  In the anti-parallel direction, two electron populations emerge, with energies of 1 keV and 50 eV labeled (1) and (2) respectively. This observation aligns with the opening of a newly reconnected closed field line, specifically in the southern region, which manifests as a dusk-wing filament. This dusk wing filament is 'younger' compared to the filaments observed in Fig.\ref{fig:alfvenfilament}f3. Furthermore, the energy distributions in Figs.\ref{fig:DualReconnection}f3-f4 (and respective 1D cuts g3-g4) represent regions with closed field lines. The eVDF corresponding to g3-f3) reveals four populations consistent with dual-reconnected field lines, where variations in bi-directional energetic flux intensities possibly suggest differing strengths of reconnection at the two sites. The eVDF in f3-g3 represents newly-reconnected closed field line, however the reconnection northward/dawnward (source of energetic anti-parallel electrons) of MMS is not as intense as the reconnection occurring southward/duskward (source of energetic parallel electrons) of MMS. The eVDF in f4-g4 also indicates a dual-reconnected field line, but with very depleted solar wind populations \textit{i.e.} populations (2) and (3), suggesting an aged dual-reconnected field line. And for this dual reconnected field line the reconnection northward/dawnward of MMS is intense compared to the southward/duskward of MMS. Furthermore, the solar wind population labeled (3) in g3 shows signatures of depletion compared to population (3) in g1 suggesting the fieldline in g3 is older than g1 i.e age of $f1 < f3 < f4$.

\section{Summary}
This work pioneers electron-distribution characterization of magnetic topologies in the interaction between sub-Alfv\'{e}nic solar wind and the Alfv\'{e}n wing magnetosphere. The CME on April 24th, 2023, presented a rare opportunity to investigate the transformation of Earth's magnetosphere into Alfv\'{e}n wings. This CME brought a magnetic cloud with low-density resulting in a long duration of sub-Alfv\'{e}nic solar wind, causing the magnetotail to split into dusk and dawn Alfv\'{e}n wings \cite{beedle_field-aligned_2024,chen_earths_2024,chen_interplanetary_2024}. During this event, MMS was positioned in the pre-noon sector. In this study, we utilized electrons as tracers  to examine the dynamic changes in magnetic topology during the Alfv\'{e}n wing transformation. 

The regions with distinct magnetic topologies include the unshocked solar wind, dusk and dawn Alfv\'{e}n wing flux tubes, and newly-closed field line regions.
The solar wind electrons exhibit a bipolar signature in $f_{||}-f_{anti,||}$ and the corresponding energy distribution functions show strahl electrons in $v_{||}<0$. The dusk (dawn) wing filaments show signatures of energetic anti-parallel (parallel) electrons which stream along these field lines. Electrons in closed field-line regions formed when the draped IMF reconnects with dusk and dawn wing field lines at the northern-dawn and southern-dusk cusps. Additionally, we identified aged dual-reconnected field lines by their pitch-angle evolved bi-directional energetic electrons, and aged dusk (dawn) wing field lines by their pitch-angle evolved energetic anti-parallel (parallel) electrons. Newly dual-reconnected field lines exhibit four electron populations, comprising low energy ($\leq 100$ eV) solar wind electrons and bi-directional energetic ($\geq 1$ keV) electrons. 


The reconnection process involves coupling multiple scales, making it difficult to identify an active reconnection process away from the diffusion region. Typically, reconnection is identified by standard signatures, such as deviations of the electron and ion bulk flow speeds from the $\mathbf{E} \times \mathbf{B}$ drift. However, these signatures cannot be used in this study because the MMS path does not pass through the diffusion region. In this study we demonstrate the signatures of active and dormant reconnection utilizing the signatures of the field-aligned energetic electrons in the VDFs. An active reconnection site north/south of MMS continuously produces parallel/anti-parallel $\sim 1$ keV electrons. We demonstrate the cessation of reconnection during Alfv\'{e}n wing formation from the absence of field-aligned electrons in the VDFs within similar topology. The study also demonstrated the importance of analyzing three-dimensional velocity distributions across all energy levels in a complex and dynamically evolving topologies to accurately confirm their identity. A similar approach can be used to study the bursty reconnection in other space plasmas like the solar atmosphere, where eruptive processes are time-limited \cite{Priest_Forbes}, and at the magnetopause of the Earth, where flux transfer events (FTEs) are signatures of a temporally and spatially limited reconnection process \cite{Russel_Elphic} and bursty bulk flows \cite{Angelopoulos}.

\section{Open Research}
The data from MMS used in this study can be accessed from the FPI and FIELDS datasets available at the MMS Science Data Center, Laboratory for Atmospheric and Space Physics (LASP), University of Colorado Boulder via \url{https://lasp.colorado.edu/mms/sdc/public/datasets/}. Additionally, data from the following MMS1 datasets were utilized:\cite{GershmanMMS_2,Gershman_MMS} and \cite{Russell}.

\acknowledgments

HG and JRS were supported by MMS Early Career grant 80NSSC21K148. RED was supported by the MMS Theory and Modeling program through NASA grant 80NSSC22K1109.

\bibliography{agusample}

\begin{thebibliography}{}

\bibitem [\protect \citeauthoryear {%
Angelopoulos%
\ \protect \BOthers {.}}{%
Angelopoulos%
\ \protect \BOthers {.}}{%
{\protect \APACyear {1992}}%
}]{%
Angelopoulos}
\APACinsertmetastar {%
Angelopoulos}%
\begin{APACrefauthors}%
Angelopoulos, V.%
, Baumjohann, W.%
, Kennel, C\BPBI F.%
, Coroniti, F\BPBI V.%
, Kivelson, M\BPBI G.%
, Pellat, R.%
\BDBL {}Paschmann, G.%
\end{APACrefauthors}%
\unskip\
\newblock
\APACrefYearMonthDay{1992}{}{}.
\newblock
{\BBOQ}\APACrefatitle {Bursty bulk flows in the inner central plasma sheet} {Bursty bulk flows in the inner central plasma sheet}.{\BBCQ}
\newblock
\APACjournalVolNumPages{Journal of Geophysical Research: Space Physics}{97}{A4}{4027-4039}.
\newblock
\begin{APACrefURL} \url{https://agupubs.onlinelibrary.wiley.com/doi/abs/10.1029/91JA02701} \end{APACrefURL}
\newblock
\begin{APACrefDOI} \doi{https://doi.org/10.1029/91JA02701} \end{APACrefDOI}
\PrintBackRefs{\CurrentBib}

\bibitem [\protect \citeauthoryear {%
Bavassano~Cattaneo%
\ \protect \BOthers {.}}{%
Bavassano~Cattaneo%
\ \protect \BOthers {.}}{%
{\protect \APACyear {2006}}%
}]{%
Bavassano_2006}
\APACinsertmetastar {%
Bavassano_2006}%
\begin{APACrefauthors}%
Bavassano~Cattaneo, M\BPBI B.%
, Marcucci, M\BPBI F.%
, Retinò, A.%
, Pallocchia, G.%
, Rème, H.%
, Dandouras, I.%
\BDBL {}Balogh, A.%
\end{APACrefauthors}%
\unskip\
\newblock
\APACrefYearMonthDay{2006}{}{}.
\newblock
{\BBOQ}\APACrefatitle {Kinetic signatures during a quasi-continuous lobe reconnection event: Cluster Ion Spectrometer (CIS) observations} {Kinetic signatures during a quasi-continuous lobe reconnection event: Cluster ion spectrometer (cis) observations}.{\BBCQ}
\newblock
\APACjournalVolNumPages{Journal of Geophysical Research: Space Physics}{111}{A9}{}.
\newblock
\begin{APACrefURL} \url{https://agupubs.onlinelibrary.wiley.com/doi/abs/10.1029/2006JA011623} \end{APACrefURL}
\newblock
\begin{APACrefDOI} \doi{https://doi.org/10.1029/2006JA011623} \end{APACrefDOI}
\PrintBackRefs{\CurrentBib}

\bibitem [\protect \citeauthoryear {%
Beedle%
\ \protect \BOthers {.}}{%
Beedle%
\ \protect \BOthers {.}}{%
{\protect \APACyear {2024}}%
}]{%
beedle_field-aligned_2024}
\APACinsertmetastar {%
beedle_field-aligned_2024}%
\begin{APACrefauthors}%
Beedle, J\BPBI M\BPBI H.%
, Chen, L\BHBI J.%
, Shuster, J\BPBI R.%
, Gurram, H.%
, Gershman, D\BPBI J.%
, Chen, Y.%
\BDBL {}Torbert, R\BPBI B.%
\end{APACrefauthors}%
\unskip\
\newblock
\APACrefYearMonthDay{2024}{{\APACmonth{02}}}{}.
\newblock
\APACrefbtitle {Field-{Aligned} {Current} {Structures} during the {Terrestrial} {Magnetosphere}'s {Transformation} into {Alfven} {Wings} and {Recovery}.} {Field-{Aligned} {Current} {Structures} during the {Terrestrial} {Magnetosphere}'s {Transformation} into {Alfven} {Wings} and {Recovery}.}
\newblock
\APACaddressPublisher{}{arXiv}.
\newblock
\begin{APACrefURL} [{2024-04-01}]\url{http://arxiv.org/abs/2402.16895} \end{APACrefURL}
\newblock
\APACrefnote{arXiv:2402.16895 [astro-ph, physics:physics]}
\PrintBackRefs{\CurrentBib}

\bibitem [\protect \citeauthoryear {%
Borovsky%
}{%
Borovsky%
}{%
{\protect \APACyear {2021}}%
}]{%
strahl}
\APACinsertmetastar {%
strahl}%
\begin{APACrefauthors}%
Borovsky, J\BPBI E.%
\end{APACrefauthors}%
\unskip\
\newblock
\APACrefYearMonthDay{2021}{}{}.
\newblock
{\BBOQ}\APACrefatitle {Exploring the Properties of the Electron Strahl at 1 AU as an Indicator of the Quality of the Magnetic Connection Between the Earth and the Sun} {Exploring the properties of the electron strahl at 1 au as an indicator of the quality of the magnetic connection between the earth and the sun}.{\BBCQ}
\newblock
\APACjournalVolNumPages{Frontiers in Astronomy and Space Sciences}{8}{}{}.
\newblock
\begin{APACrefURL} \url{https://www.frontiersin.org/articles/10.3389/fspas.2021.646443} \end{APACrefURL}
\newblock
\begin{APACrefDOI} \doi{10.3389/fspas.2021.646443} \end{APACrefDOI}
\PrintBackRefs{\CurrentBib}

\bibitem [\protect \citeauthoryear {%
Burch%
\ \protect \BOthers {.}}{%
Burch%
\ \protect \BOthers {.}}{%
{\protect \APACyear {2016}}%
}]{%
burch2016}
\APACinsertmetastar {%
burch2016}%
\begin{APACrefauthors}%
Burch, J\BPBI L.%
, Torbert, R\BPBI B.%
, Phan, T\BPBI D.%
, Chen, L\BHBI J.%
, Moore, T\BPBI E.%
, Ergun, R\BPBI E.%
\BDBL {}Chandler, M.%
\end{APACrefauthors}%
\unskip\
\newblock
\APACrefYearMonthDay{2016}{}{}.
\newblock
{\BBOQ}\APACrefatitle {Electron-scale measurements of magnetic reconnection in space} {Electron-scale measurements of magnetic reconnection in space}.{\BBCQ}
\newblock
\APACjournalVolNumPages{Science}{352}{6290}{aaf2939}.
\newblock
\begin{APACrefURL} \url{https://www.science.org/doi/abs/10.1126/science.aaf2939} \end{APACrefURL}
\newblock
\begin{APACrefDOI} \doi{10.1126/science.aaf2939} \end{APACrefDOI}
\PrintBackRefs{\CurrentBib}

\bibitem [\protect \citeauthoryear {%
Burkholder%
\ \protect \BOthers {.}}{%
Burkholder%
\ \protect \BOthers {.}}{%
{\protect \APACyear {2024}}%
}]{%
burkholder_global_2024}
\APACinsertmetastar {%
burkholder_global_2024}%
\begin{APACrefauthors}%
Burkholder, B\BPBI L.%
, Chen, L\BHBI J.%
, Sarantos, M.%
, Gershman, D\BPBI J.%
, Argall, M\BPBI R.%
, Chen, Y.%
\BDBL {}Gurram, H.%
\end{APACrefauthors}%
\unskip\
\newblock
\APACrefYearMonthDay{2024}{}{}.
\newblock
{\BBOQ}\APACrefatitle {Global {Magnetic} {Reconnection} {During} {Sustained} {Sub}-{Alfv{\'e}nic} {Solar} {Wind} {Driving}} {Global {Magnetic} {Reconnection} {During} {Sustained} {Sub}-{Alfv{\'e}nic} {Solar} {Wind} {Driving}}.{\BBCQ}
\newblock
\APACjournalVolNumPages{Geophysical Research Letters}{51}{6}{e2024GL108311}.
\newblock
\begin{APACrefURL} [{2024-04-01}]\url{https://onlinelibrary.wiley.com/doi/abs/10.1029/2024GL108311} \end{APACrefURL}
\newblock
\APACrefnote{\_eprint: https://onlinelibrary.wiley.com/doi/pdf/10.1029/2024GL108311}
\newblock
\begin{APACrefDOI} \doi{10.1029/2024GL108311} \end{APACrefDOI}
\PrintBackRefs{\CurrentBib}

\bibitem [\protect \citeauthoryear {%
Chané%
, Saur%
, Neubauer%
, Raeder%
\BCBL {}\ \BBA {} Poedts%
}{%
Chané%
\ \protect \BOthers {.}}{%
{\protect \APACyear {2012}}%
}]{%
chane_2012}
\APACinsertmetastar {%
chane_2012}%
\begin{APACrefauthors}%
Chané, E.%
, Saur, J.%
, Neubauer, F\BPBI M.%
, Raeder, J.%
\BCBL {}\ \BBA {} Poedts, S.%
\end{APACrefauthors}%
\unskip\
\newblock
\APACrefYearMonthDay{2012}{}{}.
\newblock
{\BBOQ}\APACrefatitle {Observational evidence of Alfvén wings at the Earth} {Observational evidence of alfvén wings at the earth}.{\BBCQ}
\newblock
\APACjournalVolNumPages{Journal of Geophysical Research: Space Physics}{117}{A9}{}.
\newblock
\begin{APACrefURL} \url{https://agupubs.onlinelibrary.wiley.com/doi/abs/10.1029/2012JA017628} \end{APACrefURL}
\newblock
\begin{APACrefDOI} \doi{https://doi.org/10.1029/2012JA017628} \end{APACrefDOI}
\PrintBackRefs{\CurrentBib}

\bibitem [\protect \citeauthoryear {%
L\BHBI J.~Chen%
\ \protect \BOthers {.}}{%
L\BHBI J.~Chen%
\ \protect \BOthers {.}}{%
{\protect \APACyear {2024}}%
}]{%
chen_earths_2024}
\APACinsertmetastar {%
chen_earths_2024}%
\begin{APACrefauthors}%
Chen, L\BHBI J.%
, Gershman, D.%
, Burkholder, B.%
, Chen, Y.%
, Sarantos, M.%
, Jian, L.%
\BDBL {}Burch, J.%
\end{APACrefauthors}%
\unskip\
\newblock
\APACrefYearMonthDay{2024}{{\APACmonth{03}}}{}.
\newblock
\APACrefbtitle {Earth's {Alfv}{\textbackslash}'en wings driven by the {April} 2023 {Coronal} {Mass} {Ejection}.} {Earth's {Alfv}{\textbackslash}'en wings driven by the {April} 2023 {Coronal} {Mass} {Ejection}.}
\newblock
\APACaddressPublisher{}{arXiv}.
\newblock
\begin{APACrefURL} [{2024-04-01}]\url{http://arxiv.org/abs/2402.08091} \end{APACrefURL}
\newblock
\APACrefnote{arXiv:2402.08091 [astro-ph, physics:physics]}
\PrintBackRefs{\CurrentBib}

\bibitem [\protect \citeauthoryear {%
Y.~Chen%
, Dong%
, Chen%
, Sarantos%
\BCBL {}\ \BBA {} Burkholder%
}{%
Y.~Chen%
\ \protect \BOthers {.}}{%
{\protect \APACyear {2024}}%
}]{%
chen_interplanetary_2024}
\APACinsertmetastar {%
chen_interplanetary_2024}%
\begin{APACrefauthors}%
Chen, Y.%
, Dong, C.%
, Chen, L\BHBI J.%
, Sarantos, M.%
\BCBL {}\ \BBA {} Burkholder, B\BPBI L.%
\end{APACrefauthors}%
\unskip\
\newblock
\APACrefYearMonthDay{2024}{{\APACmonth{02}}}{}.
\newblock
\APACrefbtitle {Interplanetary magnetic field \${B}\_y\$ controlled {Alfv}{\textbackslash}'\{e\}n wings at {Earth} during encounter of a coronal mass ejection.} {Interplanetary magnetic field \${B}\_y\$ controlled {Alfv}{\textbackslash}'\{e\}n wings at {Earth} during encounter of a coronal mass ejection.}
\newblock
\APACaddressPublisher{}{arXiv}.
\newblock
\begin{APACrefURL} [{2024-04-01}]\url{http://arxiv.org/abs/2402.04282} \end{APACrefURL}
\newblock
\APACrefnote{arXiv:2402.04282 [physics]}
\newblock
\begin{APACrefDOI} \doi{10.48550/arXiv.2402.04282} \end{APACrefDOI}
\PrintBackRefs{\CurrentBib}

\bibitem [\protect \citeauthoryear {%
Gershman%
\ \protect \BOthers {.}}{%
Gershman%
\ \protect \BOthers {.}}{%
{\protect \APACyear {2022}}%
{\protect \APACexlab {{\protect \BCnt {1}}}}}]{%
GershmanMMS_2}
\APACinsertmetastar {%
GershmanMMS_2}%
\begin{APACrefauthors}%
Gershman, D\BPBI J.%
, Giles, B\BPBI L.%
, Pollock, C\BPBI J.%
, Moore, T\BPBI E.%
, Kreisler, S.%
\BCBL {}\ \BBA {} Burch, J\BPBI L.%
\end{APACrefauthors}%
\unskip\
\newblock
\APACrefYearMonthDay{2022{\protect \BCnt {1}}}{}{}.
\newblock
{\BBOQ}\APACrefatitle {MMS 1 Fast Plasma Investigation, Dual Ion Spectrometer (FPI, DIS) Distribution Moments, Level 2 (L2), Fast Mode, 4.5 s Data} {Mms 1 fast plasma investigation, dual ion spectrometer (fpi, dis) distribution moments, level 2 (l2), fast mode, 4.5 s data}.{\BBCQ}
\newblock
\begin{APACrefURL} \url{http://doi.org/10.48322/PJ0N-M695} \end{APACrefURL}
\PrintBackRefs{\CurrentBib}

\bibitem [\protect \citeauthoryear {%
Gershman%
\ \protect \BOthers {.}}{%
Gershman%
\ \protect \BOthers {.}}{%
{\protect \APACyear {2022}}%
{\protect \APACexlab {{\protect \BCnt {2}}}}}]{%
Gershman_MMS}
\APACinsertmetastar {%
Gershman_MMS}%
\begin{APACrefauthors}%
Gershman, D\BPBI J.%
, Giles, B\BPBI L.%
, Pollock, C\BPBI J.%
, Moore, T\BPBI E.%
, Kreisler, S.%
\BCBL {}\ \BBA {} Burch, J\BPBI L.%
\end{APACrefauthors}%
\unskip\
\newblock
\APACrefYearMonthDay{2022{\protect \BCnt {2}}}{}{}.
\newblock
{\BBOQ}\APACrefatitle {MMS 1 Fast Plasma Investigation, Dual Ion Spectrometer (FPI, DIS) Instrument Distributions, Level 2 (L2), Burst Mode, 0.15 s Data} {Mms 1 fast plasma investigation, dual ion spectrometer (fpi, dis) instrument distributions, level 2 (l2), burst mode, 0.15 s data}.{\BBCQ}
\newblock
\begin{APACrefURL} \url{https://doi.org/10.48322/dq1y-nf73} \end{APACrefURL}
\PrintBackRefs{\CurrentBib}

\bibitem [\protect \citeauthoryear {%
Lavraud%
, Fedorov%
\BCBL {}\ \protect \BOthers {.}}{%
Lavraud%
, Fedorov%
\BCBL {}\ \protect \BOthers {.}}{%
{\protect \APACyear {2005}}%
}]{%
Lavraud_2005b}
\APACinsertmetastar {%
Lavraud_2005b}%
\begin{APACrefauthors}%
Lavraud, B.%
, Fedorov, A.%
, Budnik, E.%
, Thomsen, M\BPBI F.%
, Grigoriev, A.%
, Cargill, P\BPBI J.%
\BDBL {}Balogh, A.%
\end{APACrefauthors}%
\unskip\
\newblock
\APACrefYearMonthDay{2005}{}{}.
\newblock
{\BBOQ}\APACrefatitle {High-altitude cusp flow dependence on IMF orientation: A 3-year Cluster statistical study} {High-altitude cusp flow dependence on imf orientation: A 3-year cluster statistical study}.{\BBCQ}
\newblock
\APACjournalVolNumPages{Journal of Geophysical Research: Space Physics}{110}{A2}{}.
\newblock
\begin{APACrefURL} \url{https://agupubs.onlinelibrary.wiley.com/doi/abs/10.1029/2004JA010804} \end{APACrefURL}
\newblock
\begin{APACrefDOI} \doi{https://doi.org/10.1029/2004JA010804} \end{APACrefDOI}
\PrintBackRefs{\CurrentBib}

\bibitem [\protect \citeauthoryear {%
Lavraud%
, Thomsen%
\BCBL {}\ \protect \BOthers {.}}{%
Lavraud%
, Thomsen%
\BCBL {}\ \protect \BOthers {.}}{%
{\protect \APACyear {2005}}%
}]{%
Lavraud_2005a}
\APACinsertmetastar {%
Lavraud_2005a}%
\begin{APACrefauthors}%
Lavraud, B.%
, Thomsen, M\BPBI F.%
, Taylor, M\BPBI G\BPBI G\BPBI T.%
, Wang, Y\BPBI L.%
, Phan, T\BPBI D.%
, Schwartz, S\BPBI J.%
\BDBL {}Balogh, A.%
\end{APACrefauthors}%
\unskip\
\newblock
\APACrefYearMonthDay{2005}{}{}.
\newblock
{\BBOQ}\APACrefatitle {Characteristics of the magnetosheath electron boundary layer under northward interplanetary magnetic field: Implications for high-latitude reconnection} {Characteristics of the magnetosheath electron boundary layer under northward interplanetary magnetic field: Implications for high-latitude reconnection}.{\BBCQ}
\newblock
\APACjournalVolNumPages{Journal of Geophysical Research: Space Physics}{110}{A6}{}.
\newblock
\begin{APACrefURL} \url{https://agupubs.onlinelibrary.wiley.com/doi/abs/10.1029/2004JA010808} \end{APACrefURL}
\newblock
\begin{APACrefDOI} \doi{https://doi.org/10.1029/2004JA010808} \end{APACrefDOI}
\PrintBackRefs{\CurrentBib}

\bibitem [\protect \citeauthoryear {%
Onsager%
, Scudder%
, Lockwood%
\BCBL {}\ \BBA {} Russell%
}{%
Onsager%
\ \protect \BOthers {.}}{%
{\protect \APACyear {2001}}%
}]{%
Onsager_2001}
\APACinsertmetastar {%
Onsager_2001}%
\begin{APACrefauthors}%
Onsager, T\BPBI G.%
, Scudder, J\BPBI D.%
, Lockwood, M.%
\BCBL {}\ \BBA {} Russell, C\BPBI T.%
\end{APACrefauthors}%
\unskip\
\newblock
\APACrefYearMonthDay{2001}{}{}.
\newblock
{\BBOQ}\APACrefatitle {Reconnection at the high-latitude magnetopause during northward interplanetary magnetic field conditions} {Reconnection at the high-latitude magnetopause during northward interplanetary magnetic field conditions}.{\BBCQ}
\newblock
\APACjournalVolNumPages{Journal of Geophysical Research: Space Physics}{106}{A11}{25467-25488}.
\newblock
\begin{APACrefURL} \url{https://agupubs.onlinelibrary.wiley.com/doi/abs/10.1029/2000JA000444} \end{APACrefURL}
\newblock
\begin{APACrefDOI} \doi{https://doi.org/10.1029/2000JA000444} \end{APACrefDOI}
\PrintBackRefs{\CurrentBib}

\bibitem [\protect \citeauthoryear {%
Pollock%
\ \protect \BOthers {.}}{%
Pollock%
\ \protect \BOthers {.}}{%
{\protect \APACyear {2016}}%
}]{%
FPI}
\APACinsertmetastar {%
FPI}%
\begin{APACrefauthors}%
Pollock, C.%
, Moore, T.%
, Jacques, A.%
, Burch, J.%
, Gliese, U.%
, Saito, Y.%
\BDBL {}Zeuch, M.%
\end{APACrefauthors}%
\unskip\
\newblock
\APACrefYearMonthDay{2016}{}{}.
\newblock
{\BBOQ}\APACrefatitle {Fast Plasma Investigation for Magnetospheric Multiscale} {Fast plasma investigation for magnetospheric multiscale}.{\BBCQ}
\newblock
\APACjournalVolNumPages{Space Science Reviews}{199}{1}{331--406}.
\newblock
\begin{APACrefURL} \url{https://doi.org/10.1007/s11214-016-0245-4} \end{APACrefURL}
\newblock
\begin{APACrefDOI} \doi{10.1007/s11214-016-0245-4} \end{APACrefDOI}
\PrintBackRefs{\CurrentBib}

\bibitem [\protect \citeauthoryear {%
{Priest}%
\ \BBA {} {Forbes}%
}{%
{Priest}%
\ \BBA {} {Forbes}%
}{%
{\protect \APACyear {2002}}%
}]{%
Priest_Forbes}
\APACinsertmetastar {%
Priest_Forbes}%
\begin{APACrefauthors}%
{Priest}, E\BPBI R.%
\BCBT {}\ \BBA {} {Forbes}, T\BPBI G.%
\end{APACrefauthors}%
\unskip\
\newblock
\APACrefYearMonthDay{2002}{{\APACmonth{01}}}{}.
\newblock
{\BBOQ}\APACrefatitle {{The magnetic nature of solar flares}} {{The magnetic nature of solar flares}}.{\BBCQ}
\newblock
\APACjournalVolNumPages{The Astronomy and Astrophysics Review}{10}{4}{313-377}.
\newblock
\begin{APACrefDOI} \doi{10.1007/s001590100013} \end{APACrefDOI}
\PrintBackRefs{\CurrentBib}

\bibitem [\protect \citeauthoryear {%
Ridley%
}{%
Ridley%
}{%
{\protect \APACyear {2007}}%
}]{%
ridley_alfven_2007}
\APACinsertmetastar {%
ridley_alfven_2007}%
\begin{APACrefauthors}%
Ridley, A\BPBI J.%
\end{APACrefauthors}%
\unskip\
\newblock
\APACrefYearMonthDay{2007}{{\APACmonth{03}}}{}.
\newblock
{\BBOQ}\APACrefatitle {Alfv{\'e}n wings at {Earth}'s magnetosphere under strong interplanetary magnetic fields} {Alfv{\'e}n wings at {Earth}'s magnetosphere under strong interplanetary magnetic fields}.{\BBCQ}
\newblock
\APACjournalVolNumPages{Annales Geophysicae}{25}{2}{533--542}.
\newblock
\begin{APACrefURL} [{2023-10-22}]\url{https://angeo.copernicus.org/articles/25/533/2007/angeo-25-533-2007.html} \end{APACrefURL}
\newblock
\APACrefnote{Publisher: Copernicus GmbH}
\newblock
\begin{APACrefDOI} \doi{10.5194/angeo-25-533-2007} \end{APACrefDOI}
\PrintBackRefs{\CurrentBib}

\bibitem [\protect \citeauthoryear {%
Russell%
\ \protect \BOthers {.}}{%
Russell%
\ \protect \BOthers {.}}{%
{\protect \APACyear {2016}}%
}]{%
FGM}
\APACinsertmetastar {%
FGM}%
\begin{APACrefauthors}%
Russell, C\BPBI T.%
, Anderson, B\BPBI J.%
, Baumjohann, W.%
, Bromund, K\BPBI R.%
, Dearborn, D.%
, Fischer, D.%
\BDBL {}Richter, I.%
\end{APACrefauthors}%
\unskip\
\newblock
\APACrefYearMonthDay{2016}{}{}.
\newblock
{\BBOQ}\APACrefatitle {The Magnetospheric Multiscale Magnetometers} {The magnetospheric multiscale magnetometers}.{\BBCQ}
\newblock
\APACjournalVolNumPages{Space Science Reviews}{199}{1}{189--256}.
\newblock
\begin{APACrefURL} \url{https://doi.org/10.1007/s11214-014-0057-3} \end{APACrefURL}
\newblock
\begin{APACrefDOI} \doi{10.1007/s11214-014-0057-3} \end{APACrefDOI}
\PrintBackRefs{\CurrentBib}

\bibitem [\protect \citeauthoryear {%
Russell%
\ \BBA {} Elphic%
}{%
Russell%
\ \BBA {} Elphic%
}{%
{\protect \APACyear {1979}}%
}]{%
Russel_Elphic}
\APACinsertmetastar {%
Russel_Elphic}%
\begin{APACrefauthors}%
Russell, C\BPBI T.%
\BCBT {}\ \BBA {} Elphic, R\BPBI C.%
\end{APACrefauthors}%
\unskip\
\newblock
\APACrefYearMonthDay{1979}{}{}.
\newblock
{\BBOQ}\APACrefatitle {ISEE observations of flux transfer events at the dayside magnetopause} {Isee observations of flux transfer events at the dayside magnetopause}.{\BBCQ}
\newblock
\APACjournalVolNumPages{Geophysical Research Letters}{6}{1}{33-36}.
\newblock
\begin{APACrefURL} \url{https://agupubs.onlinelibrary.wiley.com/doi/abs/10.1029/GL006i001p00033} \end{APACrefURL}
\newblock
\begin{APACrefDOI} \doi{https://doi.org/10.1029/GL006i001p00033} \end{APACrefDOI}
\PrintBackRefs{\CurrentBib}

\bibitem [\protect \citeauthoryear {%
Russell%
\ \protect \BOthers {.}}{%
Russell%
\ \protect \BOthers {.}}{%
{\protect \APACyear {2022}}%
}]{%
Russell}
\APACinsertmetastar {%
Russell}%
\begin{APACrefauthors}%
Russell, C\BPBI T.%
, Magnes, W.%
, Wei, H.%
, Bromund, K\BPBI R.%
, Plaschke, F.%
, Fischer, D.%
\BDBL {}Burch, J\BPBI L.%
\end{APACrefauthors}%
\unskip\
\newblock
\APACrefYearMonthDay{2022}{}{}.
\newblock
{\BBOQ}\APACrefatitle {MMS 1 Flux Gate Magnetometer (FGM) DC Magnetic Field, Level 2 (L2), Burst Mode, 128 Sample/s, v4/5 Data} {Mms 1 flux gate magnetometer (fgm) dc magnetic field, level 2 (l2), burst mode, 128 sample/s, v4/5 data}.{\BBCQ}
\newblock
\begin{APACrefURL} \url{http://doi.org/10.48322/PJ0N-M695} \end{APACrefURL}
\PrintBackRefs{\CurrentBib}

\bibitem [\protect \citeauthoryear {%
Schunk%
\ \BBA {} Nagy%
}{%
Schunk%
\ \BBA {} Nagy%
}{%
{\protect \APACyear {2000}}%
}]{%
Schunk_Nagy_2000}
\APACinsertmetastar {%
Schunk_Nagy_2000}%
\begin{APACrefauthors}%
Schunk, R\BPBI W.%
\BCBT {}\ \BBA {} Nagy, A\BPBI F.%
\end{APACrefauthors}%
\unskip\
\newblock
\APACrefYear{2000}.
\newblock
\APACrefbtitle {Ionospheres: Physics, Plasma Physics, and Chemistry} {Ionospheres: Physics, plasma physics, and chemistry}.
\newblock
\APACaddressPublisher{}{Cambridge University Press}.
\PrintBackRefs{\CurrentBib}

\end{thebibliography}

\end{document}